\documentclass[]{iopart}
\usepackage{graphicx,iopams,setstack}

\usepackage[left=1in,top=1in,bottom=1in,right=1in]{geometry}

\date{\today}

\newcommand{\figwidth}{0.6\columnwidth}
\newcommand{\figwidthwide}{0.7\columnwidth}

\newcommand{\pv}{\bi{p}}
\newcommand{\qv}{\bi{q}}
\newcommand{\kv}{\bi{k}}
\newcommand{\uv}{\bi{u}}
\newcommand{\vv}{\bi{v}}

\begin{document}

\title{Enhanced Pauli blocking of light scattering in a trapped Fermi gas}

\author{B.\ Shuve\footnote{Present address: Lyman Laboratories, Harvard University, Cambridge MA 02138, USA} and J.\ H.\ Thywissen}
\address{Department of Physics and Centre for Quantum Information and Quantum Control,  University of Toronto, 60 St George, Toronto ON M5S1A7, Canada}

\date{\today}

\begin{abstract}

Pauli blocking of spontaneous emission by a single excited-state atom has been predicted to be dramatic at low temperature when the Fermi energy $E_\mathrm{F}$ exceeds the recoil energy $E_\mathrm{R}$. 
The photon scattering rate of a ground-state Fermi gas can also be suppressed by occupation of the final states accessible to a recoiling atom, however suppression is diminished by scattering events near the Fermi edge.
We analyze two new approaches to improve the visibility of Pauli blocking in a trapped Fermi gas.
Focusing the incident light to excite preferentially the high-density region of the cloud can increase the blocking signature by 14\%, and is most effective at intermediate temperature.
Spontaneous Raman scattering between imbalanced internal states can be strongly suppressed at low temperature, and is completely blocked for a final-state $E_\mathrm{F} > 4 E_\mathrm{R}$ in the high imbalance limit.

\end{abstract}



\maketitle

\section{Introduction \label{sec:intro}}

The Purcell Effect is the enhancement or reduction of scattering due to a modification of the electromagnetic density of states by a cavity \cite{purcell46}. A complementary effect can occur in an ensemble of fermions, where final states of the recoiling particle are blocked, reducing the scattering cross section. 
In semiconductors, such blocking creates a Moss-Burnstein shift \cite{mossburnstein} of the apparent band gap.
Further study of this fundamental effect has been proposed for neutral Fermi gases \cite{javanainen95,zoller98,jin98,javanainen99,busch08}, motivated by the control of trap dimensionality, the direct quantification of density, and absence of additional scattering phenomena.

Light scattering is the primary tool for detection of ultra-cold atoms \cite{ketterleVarenna}, with a few notable exceptions \cite{westbrook01}. In addition to direct imaging, light scattering has been used to probe density \cite{kleppner98}, phase coherence \cite{phillips99,aspect03}, momentum distributions \cite{ketterle99}, excitation spectra \cite{davidson02}, and superradiance \cite{phillips99b,ketterle99b,ketterle03} in quantum degenerate Bose gases. The light scattering properties of degenerate Fermi gases, by contrast, have
only recently been studied experimentally \cite{zimmermann08,jin08}, despite numerous theoretical proposals 
\cite{javanainen95,zoller98,jin98,javanainen99,ruostekoski99,hulet99,ruostekoski00,you00,zoller00,ketterle01,you01,masalas01,tosi01,gardiner07,busch08}.
In-situ optical probes could be particularly useful in exploring the physics of paired superfluids \cite{ruostekoski99,hulet99,ruostekoski00,zoller00,gardiner07}. In addition, the temperature dependence of light scattering could be exploited for thermometry.

In this work, we review two scenarios in which Pauli blocking has been considered previously, and then consider two new scenarios in which the experimental signature of blocking can be enhanced.
Whereas pioneering work treated untrapped gases \cite{javanainen95,ruostekoski99,ruostekoski00,ketterle01}, or geometries with spherical \cite{zoller98,you00,zoller00} or cylindrical \cite{zoller98,jin98,tosi01} symmetries, we treat a generalized scenario that includes finite temperature and a tri-axial trap potential. Since the high optical density of a trapped gas requires off-resonant excitation to avoid the multiple-scattering regime, we focus on scattering suppression  \cite{zoller98,jin98,tosi01,busch08} instead of line shape \cite{javanainen95,javanainen99,ruostekoski99,masalas01}. 

We develop, in sections~\ref{sec:methods} and \ref{sec:basic}, a semiclassical approach that may be applied to situations in which fully quantum calculations have proved onerous. In section~\ref{sec:basic} we show that our approach reproduces fully quantum calculations in the literature. We then calculate angle-averaged finite-temperature signatures without imposing any symmetry. Finding that suppression is rarely complete, we evaluate in section~\ref{sec:enhancement} two approaches to stronger blocking: using a focused excitation beam, and scattering between imbalanced populations. Finally, experimental prospects are discussed in section~\ref{sec:conclusion}.

\section{Methods \label{sec:methods}}

	We consider light scattering and spontaneous emission in the presence of a Fermi sea of neutral atoms. Atomic excitation is assumed to be far below saturation, and multiple scattering is assumed to be weak. The $N$ degenerate fermions are trapped in a three-dimensional harmonic trap with trap frequencies $\{ \omega_1, \omega_2, \omega_3 \} $ along three axes. When a subscript is not specified, $\omega$ refers to the geometric mean of frequencies. The single-particle Hamiltonian is
\begin{equation}
\hat{H} = \sum_j \, \frac{1}{2 m} \hat{p}_j^2 + \frac{1}{2} m \omega_j^2 \hat{q}_j^2, 
\end{equation}
where $m$ is the mass of the atom, $\hat{p}_j$ is the momentum operator in the $j$th direction, and $ \hat{q}_j$ is the position operator in the $j$th direction.

	 We treat spontaneous emission and scattering of an incident photon using a Golden Rule approach, as in \cite{zoller98,jin98}.\footnote{This approach does not treat coherent effects, which are expected to be important within the forward diffraction cone \cite{ketterleVarenna,you00}.} The reduction of the scattering rate by Pauli blocking is proportional to the reduction in the number of final states, weighted by matrix elements. This is conceptually similar to the suppression of spontaneous emission in an optical cavity where the density of electromagnetic states is reduced; here, the density of available atomic states is reduced.
	We define the relative scattering rate $S$ to be the ratio of the scattering rate with fermions to the scattering rate with Boltzmann particles, i.e., the rate without blocking effects:
			
\begin{equation} \label{eq:SQM}
	S(\kv) = \frac{ \sum_\vv \sum_\uv  n_i(\vv) \{ 1 - n_f(\uv) \} \left| \langle \uv | e^{i \kv \cdot \hat{\bi{q}}} | \vv \rangle  \right|^2 } 
	{ \sum_\vv \sum_\uv  n_i(\vv) \left| \langle \uv | e^{i \kv \cdot \hat{\bi{q}}} | \vv \rangle  \right|^2  },  
\end{equation}
	where $\kv$ is the recoil momentum of the atom, $ | \vv \rangle$ and $ | \uv \rangle$ are energy eigenstates $| v_1, v_2, v_3 \rangle$ and $| u_1, u_2, u_3 \rangle$, and $n_i$ and $n_f$ are the initial and final occupation functions, respectively%
\footnote{In Fermi gases, unlike in Bose gases, neglecting the zero point energy $\hbar \sum_j \omega_j /2$ when calculating the occupation $n_{i,f}$ leads to a fractional error in the chemical potential of $(1+\epsilon/2)(6 N \epsilon)^{-1/3}$, where eccentricity $\epsilon \equiv \omega_z/\omega_\perp$.}. %
The matrix element along a single direction can be calculated using \cite{wineland79}

\begin{equation}
\left| \langle u_j | e^{i k_j \hat{x_j}} | v_j \rangle  \right|^2 = e^{-(k_j x_0)^2} \frac{w!}{(w+\Delta)!} L^{\Delta}_w [(k_j x_{0j})^2] ^2,
\end{equation}
	where $k_j$ is the projection of $\kv$ along the $j^\mathrm{th}$ direction, $x_{0j} = \sqrt{\hbar/2 M \omega_j}$ is the ground state width, $w=\mathrm{min}(u_j,v_j)$, $\Delta = |u_j - v_j|$, and $L^\alpha_n (z)$ is the generalized Laguerre polynomial.
	
Due to the computational demands of a six-dimensional sum, (\ref{eq:SQM}) is more easily calculated in the case of a spherically symmetric trap, where three of the sums can be eliminated \cite{zoller98}. A similar approach applied to the case of cylindrical symmetry can reduce six sums to five. Without these symmetries, our desktop computer did not have the resources to calculate $S$ with experimentally realistic parameters (e.g., $N=10^6$ atoms, and trap frequencies of $\vec{\omega} / 2 \pi = \{ 500, 800, 30 \} $~Hz) at finite temperature. Indeed, no such calculation has been published.

If many states are occupied along all three harmonic axes, a semiclassical integral may capture the important physics of the problem. One might also expect a local density approach to be appropriate since light scattering predominantly probes high-momentum properties that depend upon local density fluctuations in the gas. Starting from the semiclassical phase space element \cite{kleppner87}
\( dN = h^{-3} n(\epsilon) d^3\pv \, d^3\qv \), where $n(\epsilon)$ is the quantum statistical occupation function, the relative scattering rate is

	\begin{equation} \label{eq:SSC}
	S(\kv) = \frac{ h^{-3} \int \int d^3\pv \, d^3\qv \, n_i(\pv,\qv) \{ 1 - n_f(\pv+\kv,\qv) \} } {h^{-3} \int \int d^3\pv \, d^3\qv \, n_i(\pv,\qv) } .
	\end{equation}
This local density approach neglects the energy quantization scale set by the level spacing, and thereby allows us to rescale the problem into an isotropic one, even though no such symmetry is manifest in the trap geometry. The symmetry is broken only by the direction of the momentum recoil due to scattering.\footnote{For instance, illumination of a cigar-shaped cloud along one of its two radial axes would break all rotational symmetries.} A semiclassical approximation was also used in \cite{masalas01} to discuss line shape, and to treat the uniform gas in \cite{ketterle01}. Here we extend these treatments to find the light scattering properties of a trapped gas.

\begin{figure}[b!]
\centerline{\includegraphics[width=\figwidth]{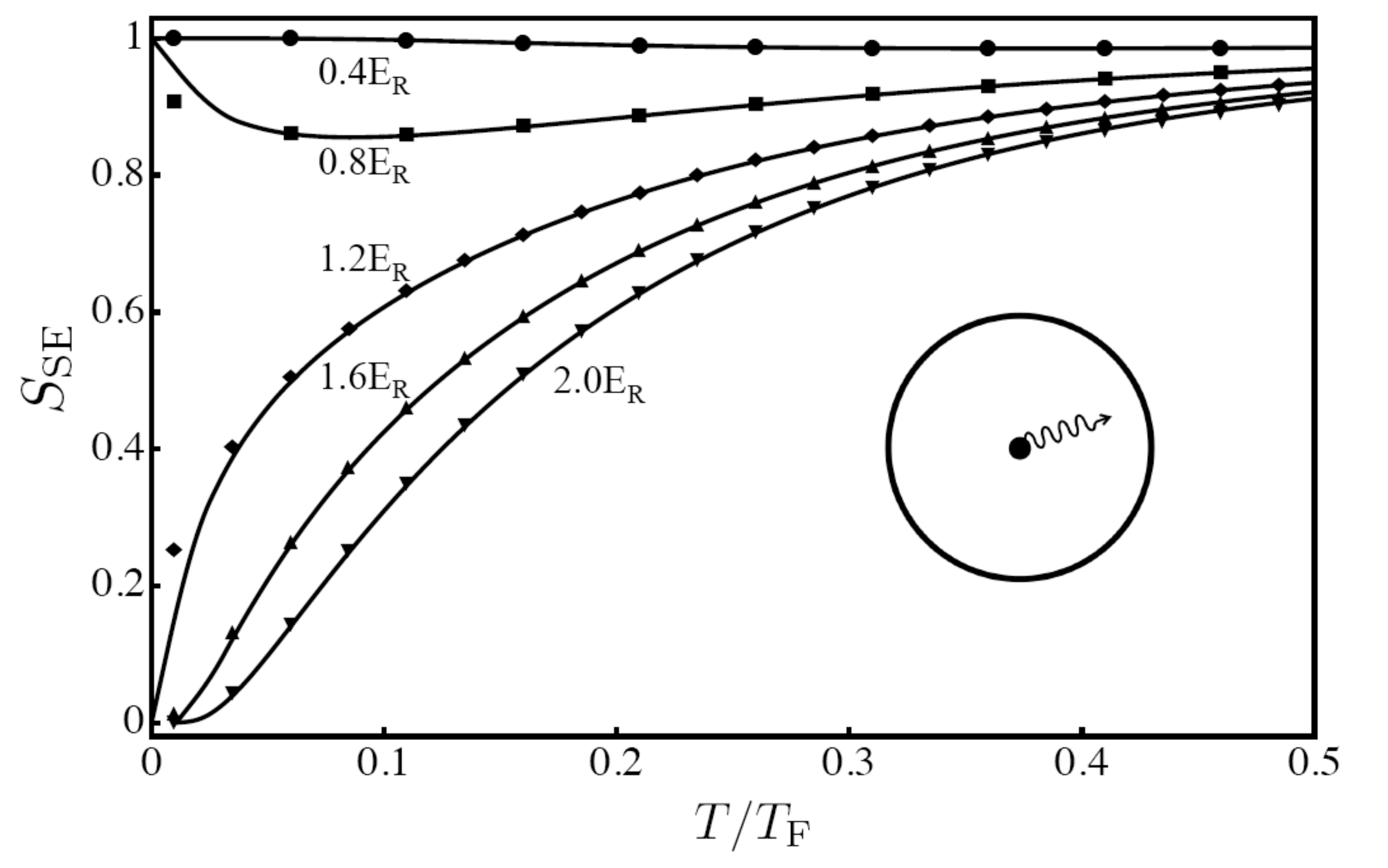}}%
\caption{Comparison of quantum (points) and semiclassical (lines) calculation of spontaneous emission blocking. Suppression is absent when $S \rightarrow 1$ and complete when $S=0$.  For comparison to figure~1 in \cite{zoller98}, the quantum calculations assume $E_\mathrm{R}= 25 \hbar \omega$ and spherical symmetry, however these assumptions are not necessary for the semiclassical calculation. From top to bottom, $E_\mathrm{F}/E_\mathrm{R}$ = 0.4, 0.8, 1.2, 1.6, 2.0. 
The inset ``cartoon'' is a schematic representation of the process under consideration: an atom in the midst of a Fermi sea emitting a single photon.
\label{Fig1}}
\end{figure}

For large atom numbers and moderate trap anisotropies, we show (in section~\ref{sec:basic}) that (\ref{eq:SSC}) is in excellent agreement with published calculations based on (\ref{eq:SQM}). Furthermore, the simplicity of the semiclassical method allows us to include angular averaging across the scattered photon momentum, to evaluate both finite and zero temperatures, and (in section~\ref{sec:enhancement}) to consider more complex scenarios.

\section{Signatures of Pauli Blocking \label{sec:basic}}

Two scenarios for Pauli blocking have been considered in previous work. The first is spontaneous emission (``SE") of a single excited-state atom in the midst of a Fermi sea. The second is light scattering (``LS") off a large ground-state ensemble. For reasons discussed further in section~\ref{sec:SE} and in section \ref{sec:conclusion}, the LS scenario is experimentally more feasible, and is the focus of our work. However, the SE case is the simplest scenario that elucidates the blocking effect under discussion. Our treatment can also be compared directly to \cite{zoller98}, which presents a fully quantum calculation of the same scenario.

\subsection{Spontaneous emission of a single excited atom\label{sec:SE}}
\begin{figure}[t] 
\centerline{\includegraphics[width=\figwidth]{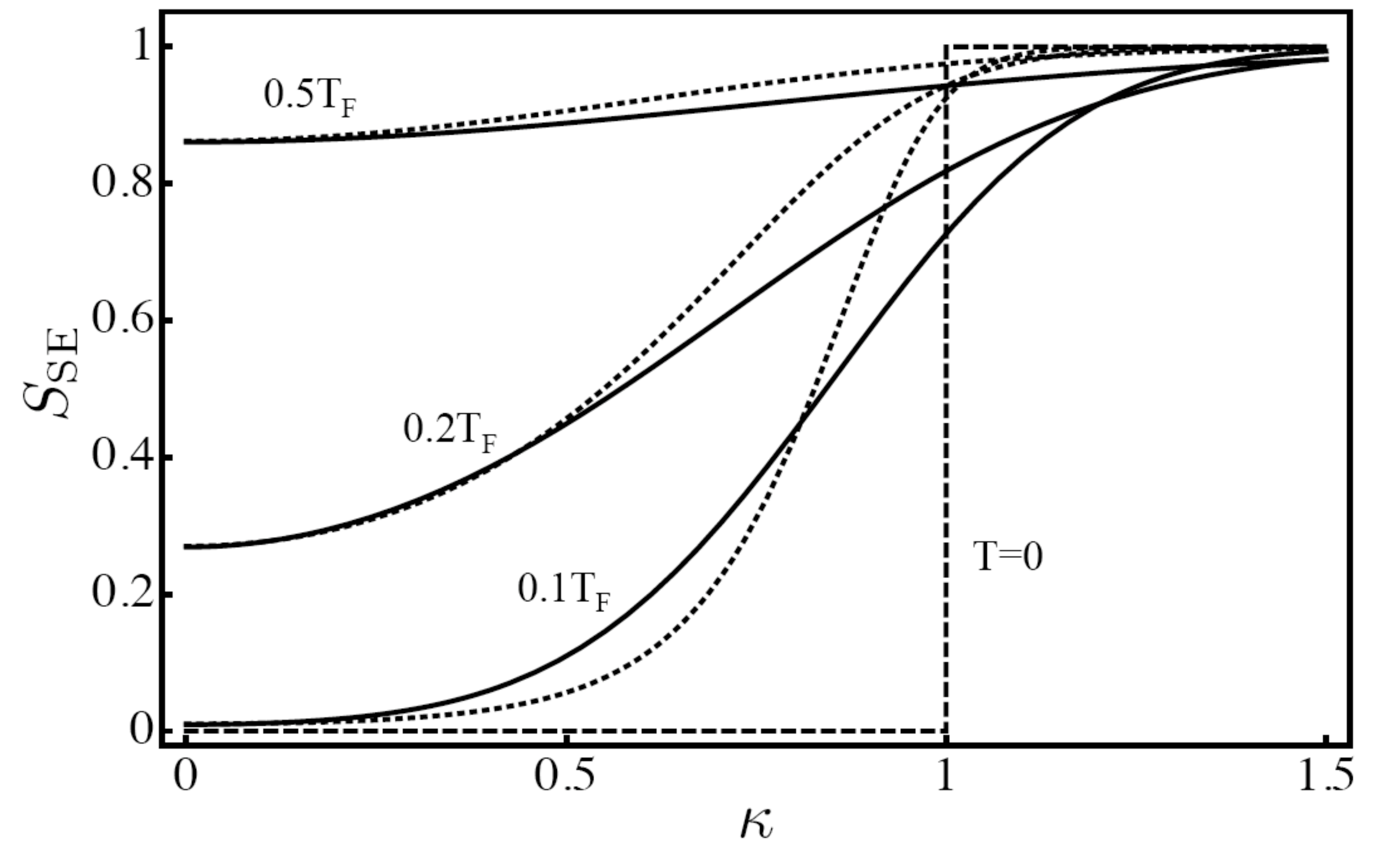}}%
\caption{Spontaneous emission rate $S_\mathrm{SE}$ versus momentum kick $\kappa =(E_\mathrm{R}/E_\mathrm{F})^{1/2}$. Solid lines are numerical integrals as defined in (\ref{eq:SSE1}), for $k_\mathrm{B} T/E_\mathrm{F} = 0.5$, 0.2, and 0.1, as labelled. Short-dashed lines are the constant kick approximate form of (\ref {eq:avkick}); long-dashed lines are the zero temperature limit (\ref{eq:SEzeroT}).
\label{Fig2}} 
\end{figure}

	For a single atom in the excited state decaying into a Fermi sea of $N$ atoms, 
	 \begin{equation}
	 n_i(\epsilon)  = (\beta \hbar \omega)^3 \exp{(-\beta \epsilon)}
	 \qquad\mbox{and}\qquad
	 n_f(\epsilon) = [ 1 + z^{-1} \exp{\beta \epsilon} ]^{-1},  \label{eq:ninfSE}
	 \end{equation}
	 where $\beta^{-1} = k_\mathrm{B} T$, $k_\mathrm{B}$ is the Boltzmann constant, $T$ is the temperature, and $z$ is the fugacity of the Fermi gas. This $n_i$ is normalized by integration over $h^{-3} d^3\pv \, d^3 \qv$.

\begin{figure}[b] 
\centerline{\includegraphics[width=\figwidth]{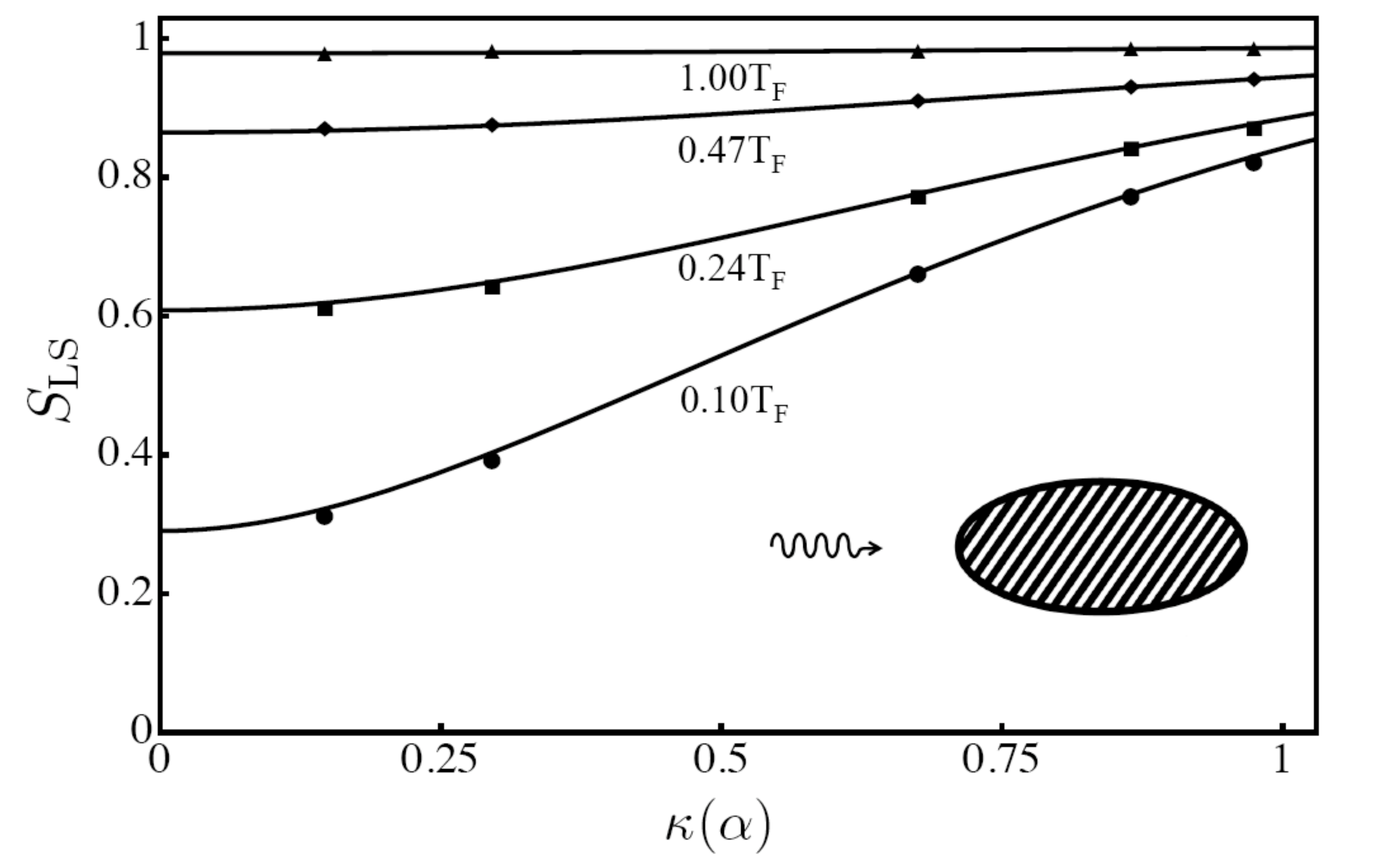}}%
\caption{Scattering rate $S_\mathrm{LS}$ versus momentum kick $\kappa(\alpha)$ comparing fully quantum (points) and semiclassical (lines) calculations, from Eqs.~(\ref{eq:SQM}) and (\ref{eq:S2}) respectively. Temperatures shown are $k_\mathrm{B} T/E_\mathrm{F}$ = 1.0 (triangles), 0.47 (circles), 0.24 (squares), and 0.10 (inverted triangles). The fully quantum calculation is for a cylindrically symmetric geometry ($\omega_1 = \omega_2 \equiv \omega_\perp$, and incident beam along axis of symmetry), and parameters are chosen to reproduce figure~3 in \cite{jin98}: $E_\mathrm{R} = 21 \hbar \omega_\perp$ and aspect ratio is 0.1. Assumptions of symmetry, aspect ratio, and $E_\mathrm{R} / \omega_\perp$ are not used for the semiclassical calculation.  
\label{Fig3}}
\end{figure}	
	 	 Note that we assume that the excited-state atom sees the same trapping potential as the ground state, and is thermalized with the atoms in the ground state. This might be realistic for fermions with long-lived metastable states, for instance in rare earth metals \cite{takahashi07} in magic wavelength traps \cite{MagicLambda}.

	 Using distributions (\ref{eq:ninfSE}), we evaluate (\ref{eq:SSC}) using a change of variables. The coordinates are rotated such that one momentum axis is aligned with the momentum kick $\kv$, and the other five coordinates are combined into a five-space radius. We define a dimensionless momentum $\xi \equiv \sqrt{\beta E_\mathrm{R}}$ from the single-photon recoil energy $E_{R} = \hbar^2 k_L^2 / 2 m$, where $c k_L$ is the laser frequency. The normalized scattering rate is then	 
	 \begin{equation}  
	  S_\mathrm{SE}(\xi, T)  =  \frac{1}{\sqrt{\pi}} z^{-1} e^{\xi^2} \int_{-\infty}^\infty dy \,
	e^{2 \xi y} \, f_{5/2}(z e^{-(y+\xi)^2}), \label{eq:SSE1}
	 \end{equation}
	 where $f_n$ is the Fermi function%
\footnote{The Fermi integrals that appear in this section are of the form
\begin{equation}	
	\int_0^\infty a^{n-1} da \frac{1}{e^a / C + 1} = \Gamma(n) f_{n}(C),
\end{equation}		
where $f_n(C)$ is $-\mathrm{Li}_n(-C)$, and $\mathrm{Li}_{n}(C) = \sum_{j=1}^{\infty} C^j / j^n$ is the polylogarithmic function.}. %
Note that since we are using a local density method, trap frequencies and atom number now affect the scattering rate only through $E_\mathrm{F}$.

	Figures~\ref{Fig1} and \ref{Fig2} show the essence of the blocking effect: at low temperature, and when the Fermi energy is larger than the recoil energy, the scattering rate decreases. Figure~\ref{Fig1} compares our results to Busch {\em et al} \cite{zoller98} to find that the semiclassical $S_\mathrm{SE}$ agrees well with a fully quantum calculation, in this case calculated using (\ref{eq:SQM}) and assuming spherical symmetry.

	Figure~\ref{Fig2} shows that complete suppression of spontaneous emission can occur when the Fermi energy exceeds the recoil energy. (By contrast, suppression is only complete for infinite Fermi energy when the set of initial states is expanded to a full Fermi sea, as discussed in the next section.) At zero temperature, (\ref{eq:SSE1}) takes the simple form
\begin{equation} \label{eq:SEzeroT}
	 S_\mathrm{SE}(\kappa) =  \Theta (\kappa-1), 
\end{equation}
where $\Theta$ is the unit step function, and we use a dimensionless momentum $\kappa \equiv (E_\mathrm{R} / E_\mathrm{F})^{1/2}$ since $\xi$ is ill-defined at zero temperature. The abruptness of this step is the only discrepancy between the quantum and semiclassical calculation, as shown at the left-most points in figure~\ref{Fig1}. Interesting non-isotropic effects have been predicted in this limit \cite{zoller98}, with possible application to directional single-photon sources \cite{busch08}. Wave-function features appear when $k_\mathrm{B} T$ is comparable to the level spacing $\hbar \omega_j$. However, even at the lowest observed temperatures of approximately $0.03 E_\mathrm{F}/k_\mathrm{B}$ \cite{KetterleImbalance,HuletImbalance}, current experiments remain in the semiclassical regime.

	Figure~\ref{Fig2} shows that the emission rate $S_\mathrm{SE}(\kappa,T\geq 0.5 T_\mathrm{F}) \geq 0.86$, confirming that blocking is not dramatic in the non-degenerate regime. However for the weak effects observed at high temperature, one can use a series expansion of $f_{5/2}$:
\begin{equation} \label{eq:SEhighT}
	 S_\mathrm{SE}(\xi,T) \longrightarrow 1 - e^{-\xi^2} \sum_{n=2}^\infty \frac{(-1)^n z^n}{n^3} e^{\xi^2/n}.
 \end{equation}
This expression converges for $z<1$, i.e., $k_\mathrm{B} T /E_\mathrm{F} > 0.57$. 

An approximate form valid for all temperatures can be developed by neglecting the directionality of the momentum kick.  Given an initial atomic momentum $\pv$, the average energy transferred by a kick is simply $E_\mathrm{R}$ when averaged over a uniform distribution of atomic momenta. Using this energy difference, we can fully integrate 
$S$:%
%
\begin{equation} \label{eq:avkick}
	S_\mathrm{SE}(\xi,T) \approx z^{-1} e^{\xi^2}  \, f_{3}(z \, e^{-\xi^2}).
	\end{equation}

	Figure~\ref{Fig2} compares the approximations (\ref{eq:SEzeroT}) and (\ref{eq:avkick}) to numerical integration of $S_\mathrm{SE}$. The average-kick approximation (\ref{eq:avkick}) underestimates the blurring of the step function at finite temperatures, but is a reasonable estimate at the 20\% level and even better at low $\kappa$. In all cases, suppression is stronger for lower recoil momentum (or higher Fermi energies), since final states fall closer to the centre of the Fermi sea.

\subsection{Light scattering from a large ensemble \label{sec:scatt}}
\begin{figure}[t] 
\centerline{\includegraphics[width=\figwidth]{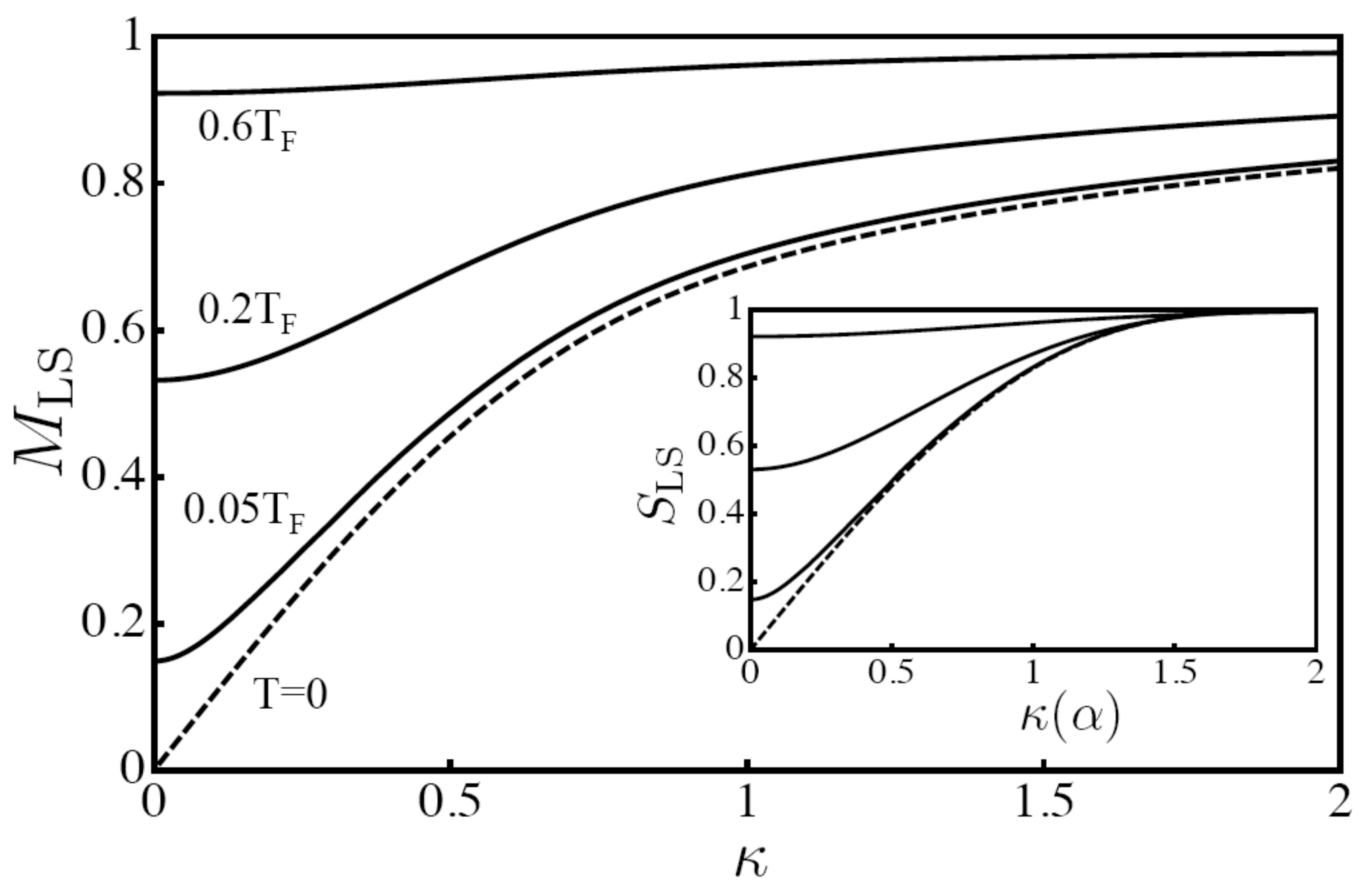}}%
\caption{Angle-averaged scattering rate $M_\mathrm{LS}$ versus average kick $\kappa$ for a dipole emission pattern. Thick lines are numerical integrals, for $k_\mathrm{B} T/E_\mathrm{F}$  = 0.6, 0.2, and 0.05. The dashed line is the zero temperature limit. {\bf Inset:} Scattering rate $S_\mathrm{LS}$ versus momentum kick $\kappa(\alpha)$, for the same temperatures.  
\label{Fig4}}
\end{figure}

	We now consider $N$ polarized fermions in a single ground state, recoiling under the net momentum $\kv$ of an incident and a scattered photon. In the perturbative limit, we ignore the disturbance of removing an atom from the distribution, and use the same initial and final state
\begin{equation}	
n_i(\epsilon)  =  n_f(\epsilon) = [ 1 + z^{-1} \exp{\beta \epsilon} ]^{-1}.
\label{eq:ninfLS}
\end{equation}
	The initial states have an energetic range that is determined both by temperature and Fermi pressure, unlike the case of a single excited-state atom. This makes it easier to scatter out of the Fermi sea, and reduces the net blocking effect. 
	
	Integration of (\ref{eq:SSC}) using (\ref{eq:ninfLS}) yields a normalized scattering rate
\begin{equation} \label{eq:S2} 
	S_\mathrm{LS}(\xi(\alpha),T) = 1 - \frac{8}{\pi (\beta E_\mathrm{F})^3} 
	\int_0^\infty a^{3/2}\, da \int_{-\infty}^\infty dy \,
	\frac{1}{1 + z^{-1} e^{a + y^2}} \frac{1}{1 + z^{-1} e^{a + (y + \xi(\alpha))^2} },
\end{equation}
where the results now depend on the angle $\alpha$ between the incident and scattered photon. The rescaled recoil momenta are $\xi(\alpha) = 2 \xi \sin^2(\alpha/2)$ and $\kappa(\alpha) = 2 \kappa \sin^2(\alpha/2)$, and we maintain the previously defined quantities $\xi^2 = \beta E_\mathrm{R}$ and $\kappa^2 = E_\mathrm{R}/E_\mathrm{F}$, written without an angle argument. Figure~\ref{Fig3} shows a numerical integration of (\ref{eq:S2}) at various temperatures. Points show a reproduction of quantum mechanical calculations assuming cylindrical symmetry, for values chosen in \cite{jin98}. The agreement is excellent. Unlike the SE case, there is no disagreement between quantum and semiclassical calculations at low temperature so long as $E_\mathrm{F} \gg \hbar \omega_j$ in all directions.

Since the excitation and decay both contribute a momentum kick, an angle-resolved experiment would observe $S$ directly \cite{jin98}. If an experiment measures the {\em total} scattering rate (see discussion in section \ref{sec:conclusion}), we observe an angle-averaged suppression factor, which we call $M$:
\begin{equation} \label{eq:M}
	M(k) =   \int_0^\pi S(2 k \cos^2{\frac{\alpha}{2}}) P(\alpha) d\alpha,
\end{equation}
where \( P(\alpha) = \frac{3}{8} (1+\cos^2{\alpha}) \) for a dipole emission pattern of any polarization, after averaging over the azimuthal scattering angle. Numerical evaluation of $M_\mathrm{LS}$ versus $\kappa$, no longer $\alpha$-dependent, is shown in figure~\ref{Fig4}. Comparing $M_\mathrm{LS}$ and $S_\mathrm{LS}$, shown in the inset, we see that angle averaging produces little qualitative change. The low-$\kappa$ limit is identical, but suppression continues to higher $\kappa$ in $M_\mathrm{LS}$. This is due to inclusion of forward-scattering events that produce small kicks and are easy to block. Including these scattering events is necessary for quantitative prediction of the scattering suppression.

As before, we can expand $S$ to find a high temperature expression,
\begin{equation} \label{eq:S2HighT}
	S_\mathrm{LS}(\xi(\alpha),T) \longrightarrow 1 - \frac{6}{(\beta E_\mathrm{F})^{3}} \sum _{n=2}^{\infty } \frac{(-1)^n z^n}{n^3} \sum _{\ell=1}^{n-1}
   e^{-\xi(\alpha) ^2 \ell (n-\ell)/n}, 
\end{equation}
however the series converges only for $z>1$, where $S_\mathrm{LS} > 0.91$. Keeping only the first term, \( S_\mathrm{LS}(\xi(\alpha)) \approx 1 - \frac{3}{4} (\beta E_\mathrm{F})^{-3} z^2 \exp{(-\xi(\alpha)^2 / 2)}.\) 

At zero temperature, an analytic expression for $S_\mathrm{LS}$ can be found:
\begin{equation} \label{eq:S2ZeroT}
S_\mathrm{LS}(\kappa(\alpha)) =  1-\frac{32}{5 \pi } \chi(\frac{\kappa(\alpha)}{2}) \Theta (2-\kappa(\alpha)),
\end{equation}
where 
\begin{equation} \label{eq:chi}
\chi(x) \equiv \frac{x \sqrt{1-x^2}}{48} (-8 x^4+26 x^2-33) +\frac{15}{48}  \cos ^{-1}(x).
\end{equation}
A further integral can also be done to find an expression for $M_\mathrm{LS}$ at zero temperature, and is given in the appendix. Both scattering rates are plotted as dashed lines in figure~\ref{Fig4}, and linearly approach zero as the momentum kick goes to zero.


Quantum corrections to the scattering rate can also be evaluated by considering (\ref{eq:SQM}) for various trap geometries, and comparing to the geometric insensitivity valid in the semiclassical limit of $E_\mathrm{F} \gg \hbar \omega_j$ for all $i=\{1,2,3\}$ trap frequencies. Evaluating (\ref{eq:SQM}) for a cylindrically symmetric trap with 20:1 aspect ratio, and comparing to a spherically symmetric trap, both for $E_\mathrm{F} = 11 \hbar \omega$, we find that $S_{SE}$ changes by less than 5\%, $S_{LS}$ changes by less than 3\%, and $M_{LS}$ changes by less than 2\%.

%
%
\begin{figure}[b] 
\centerline{\includegraphics[width=\figwidth]{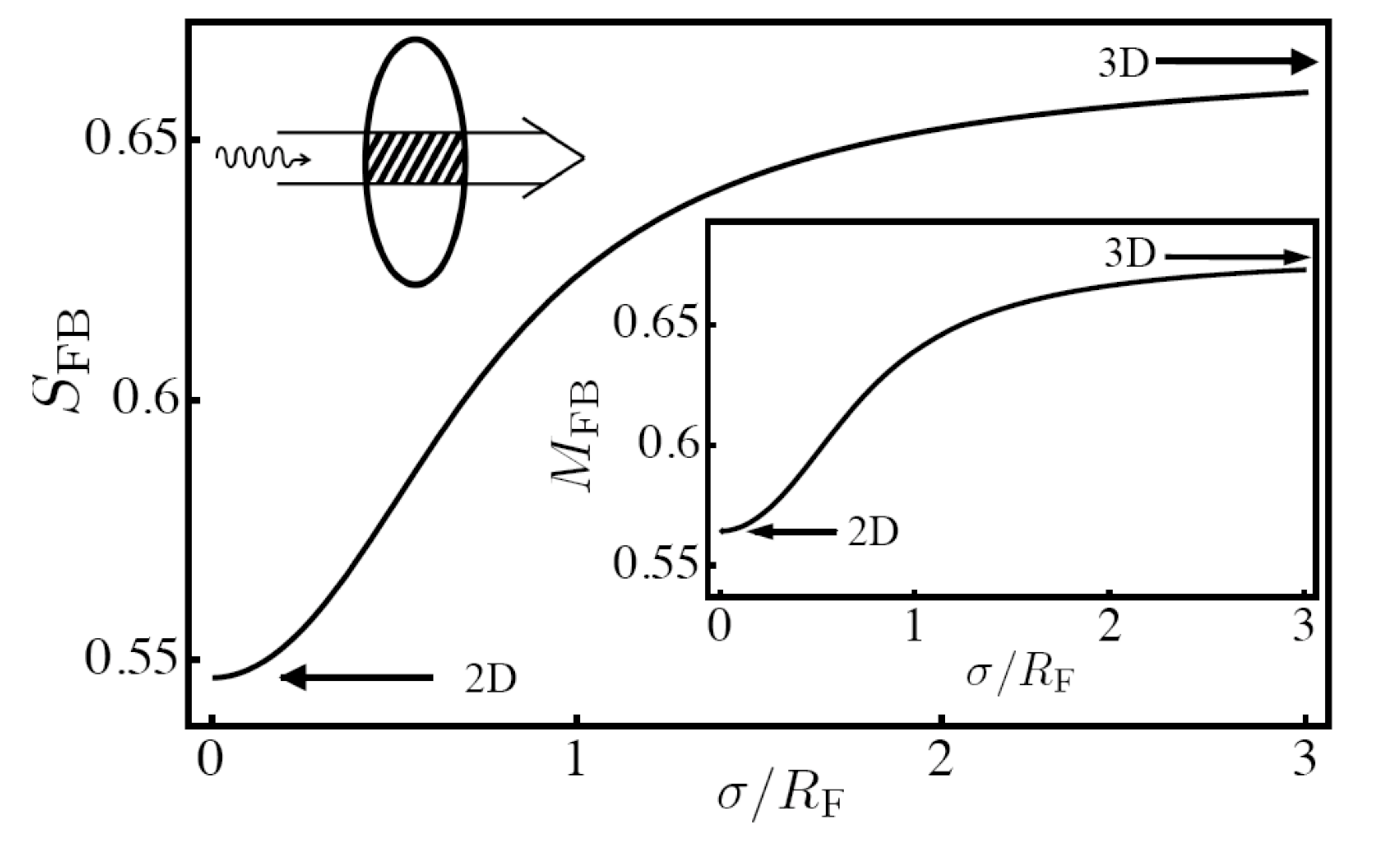}}%
\caption{Pauli blocking is enhanced when the excitation beam is focused onto the centre of the cloud. The normalized scattering rate $S_\mathrm{FB}$ is shown versus beam size $\sigma/R_F$, for $\kappa(\alpha)=0.5$ and $k_\mathrm{B} T=0.2 E_\mathrm{F}$. The arrows indicate the asymptotic limits: the three-dimensional $S_\mathrm{LS}$ at high $\sigma$, and the two-dimensional $S^\mathrm{2D}_\mathrm{LS}$ at low $\sigma$. {\bf Inset:} The same quantities are plotted with a dipole pattern angle-averaged scattering rate $M_\mathrm{FB}$.  
\label{Fig5}}
\end{figure}
\begin{figure}[t] 
\centerline{\includegraphics[width=\figwidth]{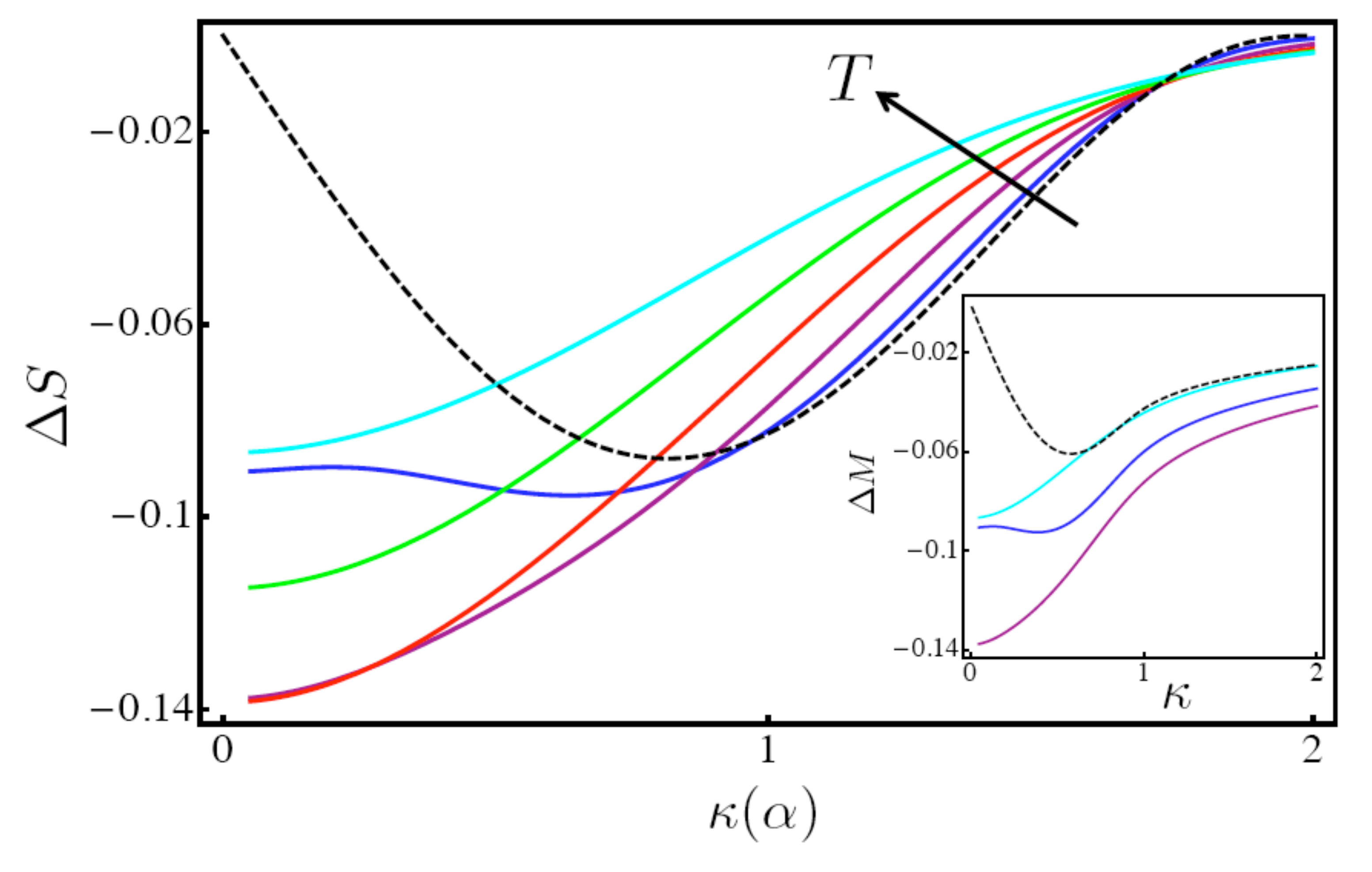}}%
\caption{The maximum possible strengthening of Pauli blocking with a focused beam, $\Delta S = S^\mathrm{2D}_\mathrm{LS} - S_\mathrm{LS}$, is shown versus momentum $\kappa(\alpha)$ for temperatures $k_\mathrm{B} T/E_\mathrm{F}$ = 0.5 (cyan), 0.4 (green), 0.3 (red), 0.2 (purple), 0.1 (blue) and zero (black dashed).  {\bf Inset:} The enhancement $\Delta M = M^\mathrm{2D}_\mathrm{LS} - M_\mathrm{LS}$ versus normalized recoil momentum $\kappa$, after averaging over a dipole angular distribution. For clarity, fewer temperatures are shown: $k_\mathrm{B} T/E_\mathrm{F}$ = 0.5 (top solid, cyan), 0.2 (bottom solid, purple), 0.1 (middle solid, blue) and zero (black dashed). Again the largest effect is at intermediate temperatures. 
\label{Fig6}}
\end{figure}

\section{Scenarios with stronger Pauli blocking \label{sec:enhancement}}

While spontaneous emission can be suppressed fully at finite momentum $\kappa$ for sufficiently low temperature, complete blocking is  possible only at $\kappa = 0$ in the case of light scattering (section~\ref{sec:scatt}). The smallest $\kappa$ reported to date is $\approx 0.6$ \cite{thomas02,thywissen06,gommers09}, so a suppression of at most $M_\mathrm{LS} \approx 0.5$ would be expected.  In this section, we explore two methods to improve the Pauli blocking signal: the use of a focused excitation beam (``FB''), and scattering between two imbalanced populations (``IP''). In both cases, we attempt to bias scattering toward events with higher local $E_\mathrm{F}$. In the FB scheme, this is done directly by selecting the spatial centre of the trap. In the IP scheme, reduced Fermi pressure in the initial state also reduces the initial kinetic energy. As is shown below, these approaches increase the overlap between accessible final states and the Fermi sea of occupation.


\subsection{Focused excitation light \label{sec:focused}}
	
	DeMarco and Jin suggest that stronger blocking might be observed when focusing the incident laser beam \cite{jin98}. Here we evaluate this scheme quantitatively. We consider excitation along a cycling transition, starting and ending in the same Fermi sea, as in section~\ref{sec:scatt}. A focused excitation beam restricts $n_i$ to its intersection with the atomic cloud. Assuming the beam propagates along $q_3$, the distribution of initial states is
\begin{equation}
	 n_i(\epsilon, q_1, q_2)  =  I(q_1, q_2) [ 1 + z^{-1} \exp{\beta \epsilon} ]^{-1},
\end{equation}
where $I(q_1, q_2)$ is a dimensionless intensity distribution of the light. The final state distribution $n_f$ remains as before, given by (\ref{eq:ninfLS}).

For simplicity we consider a cylindrically symmetric beam $I(q_1, q_2) = \exp{(-2 q_1^2/\sigma^2 - 2 q_2^2/\sigma^2)},$ where $\sigma$ is the waist of the beam. Starting from (\ref{eq:SSC}), we rescale symmetric degrees of freedom and are left with a triple integral:
%
\begin{eqnarray}
S_\mathrm{FB}(\kappa(\alpha), T, \sigma) = 
	1 - \frac{2}{\pi N_\mathrm{e}}
	\int_0^\infty  a^{\frac{1}{2}}\, da \int_0^{\infty}  x dx \int & dy \, \frac{e^{-2 (x R_F/ \sigma)^2}}{1 + z^{-1} e^{\beta E_\mathrm{F} (a + y^2+x^2)}} \nonumber\\
	 & \times \, \frac{1}{1 + z^{-1} e^{\beta E_\mathrm{F} (a + (y + \kappa(\alpha))^2+x^2)}},
\label{eq:S4}
\end{eqnarray}
where the mean number of atoms excited by the probe is
\begin{equation}
N_\mathrm{e}=\int_0^\infty a \, da \int_0^{\infty} x dx \, e^{-2 (x R_F/ \sigma)^2}
[1 + z^{-1} e^{\beta E_\mathrm{F} (a + x^2)}]^{-1},
\end{equation}
the radial Fermi radius $R_F=\sqrt{2 E_\mathrm{F}/m}/\omega_\perp$, and we have assumed $\omega_\perp \equiv \omega_1=\omega_2 $.  Figure \ref{Fig5} shows that smaller beam size enhances the suppression. Atoms are excited at the centre of the cloud, where the density is higher and thus the local $E_\mathrm{F}$ is higher. Since $E_\mathrm{R}$ is unchanged, we effectively decrease $\kappa$. 

In the large cloud (or small beam) limit, $\sigma \ll R_F$, the spatial selection of the exciting beam becomes a delta function. Since rescaled quadratic degrees of freedom are equivalent under the integral, eliminating two spatial degrees of freedom is equivalent to eliminating one spatial and one momentum degree of freedom. In other words, for a given geometric mean $\omega$, the same scattering rate is observed for a tightly focused beam on an oblate three-dimensional cloud, as would be observed for a {\it two-dimensional cloud with a uniform excitation light}. This limit is

\begin{equation} \label{eq:S2D} 
S^\mathrm{2D}_\mathrm{LS} (\xi(\alpha),T) = 
	1 - \frac{4}{\pi (\beta E_\mathrm{F})^2} 
	\int_0^\infty a^{1/2}\, da \int_{-\infty}^\infty dy \,
	\frac{1}{1 + z^{-1} e^{a + y^2}} \frac{1}{1 + z^{-1} e^{a + (y + \xi(\alpha))^2} }.
\end{equation}
Figure \ref{Fig5} shows as arrows the 3D limit (\ref{eq:S2}) and the 2D limit (\ref{eq:S2D}). 

Figure \ref{Fig6} shows the difference $\Delta S \equiv S^\mathrm{2D}_\mathrm{LS} - S_\mathrm{LS}$ (and angle-averaged $\Delta M \equiv M^\mathrm{2D}_\mathrm{LS} - M_\mathrm{LS}$) between the small- and large-beam limit. This is the maximum effect that changing beam size could have. We see that the difference is restricted to $\Delta S \lesssim 0.15$. Interestingly, the most pronounced effect occurs when $k_\mathrm{B} T/E_\mathrm{F} \approx 0.25$. At intermediate temperatures, selecting the centre of the cloud is even more important than at zero temperature, since quantum degeneracy varies across the cloud. At lower temperatures and momenta, suppression is complete for both the 2D and 3D limits, so focusing is less effective.

At zero temperature, an analytic expression can be found:
%
\begin{equation} \label{eq:S2DzT} 
	S^\mathrm{2D}_\mathrm{LS}(\kappa(\alpha)) = \frac{\Theta(2-\kappa(\alpha))}{12 \pi} \left[ \kappa(\alpha) \sqrt{4-\kappa(\alpha)^2} (\kappa(\alpha)^2 - 10) + 24 \cos^{-1}{(\kappa(\alpha)/2)} \right].
\end{equation}
This expression is shown as a dashed line in figure~\ref{Fig6}. 

A similar expression can be found for the angle-averaged $M^\mathrm{2D}_\mathrm{LS}$, and is given in the appendix. For a variety of temperatures, figures \ref{Fig5} and \ref{Fig6} show the angle-averaged results as insets. In the inset of figure~\ref{Fig6}, the dashed line shows the zero temperature difference between equations~(\ref{eq:M2zTd}) and (\ref{eq:M4zTd}). As with angle-resolved scattering, the enhancement $\Delta M$ is no more than  $0.15$, and occurs at intermediate temperature. However, because of the inclusion of low-$\alpha$ events, suppression is observed (and enhanced) for $\kappa \geq 2$.

%
\begin{figure}[b] 
\begin{center}
\centerline{\includegraphics[width=\figwidthwide]{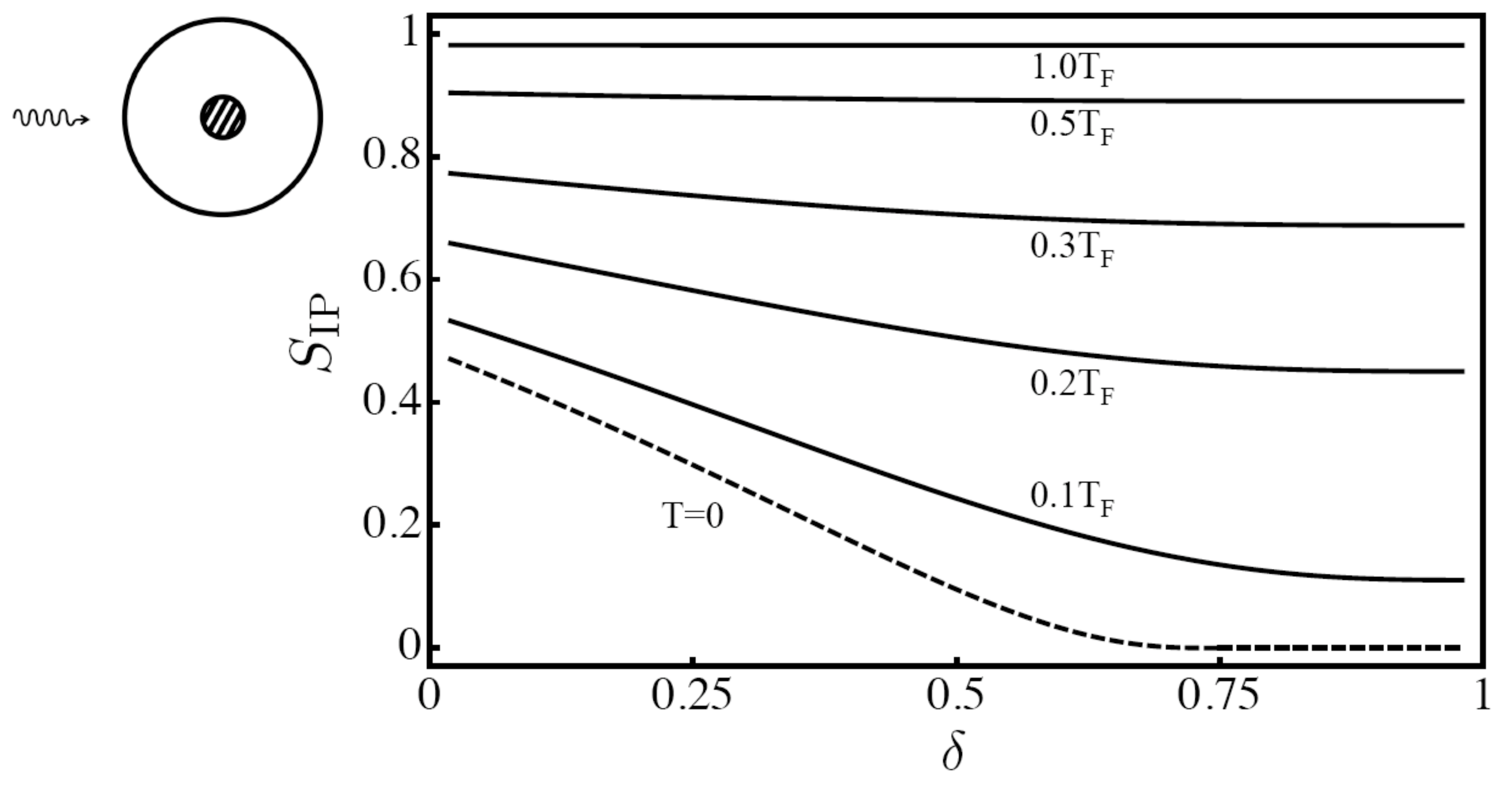}}%
\caption{Scattering rate $S_\mathrm{IP}$ versus imbalance $\delta$, at $\kappa(\alpha)=0.5$. From top to bottom, solid lines represent temperatures $k_\mathrm{B} T/E_\mathrm{F}$ = 1, 0.5, 0.3, 0.2, and 0.1 from (\ref{eq:S3}); the dashed line shows the zero temperature limit from (\ref{eq:S3zT}). Note that for Figures \ref{Fig7} and \ref{Fig8}, the final state Fermi energy has been used for dimensionless quantities, so $\kappa \equiv \sqrt{E_\mathrm{R}/E_\mathrm{Ff}}$. 
\label{Fig7}}
\end{center}
\end{figure}
\begin{figure}[t] 
\begin{center}
\centerline{\includegraphics[width=\figwidth]{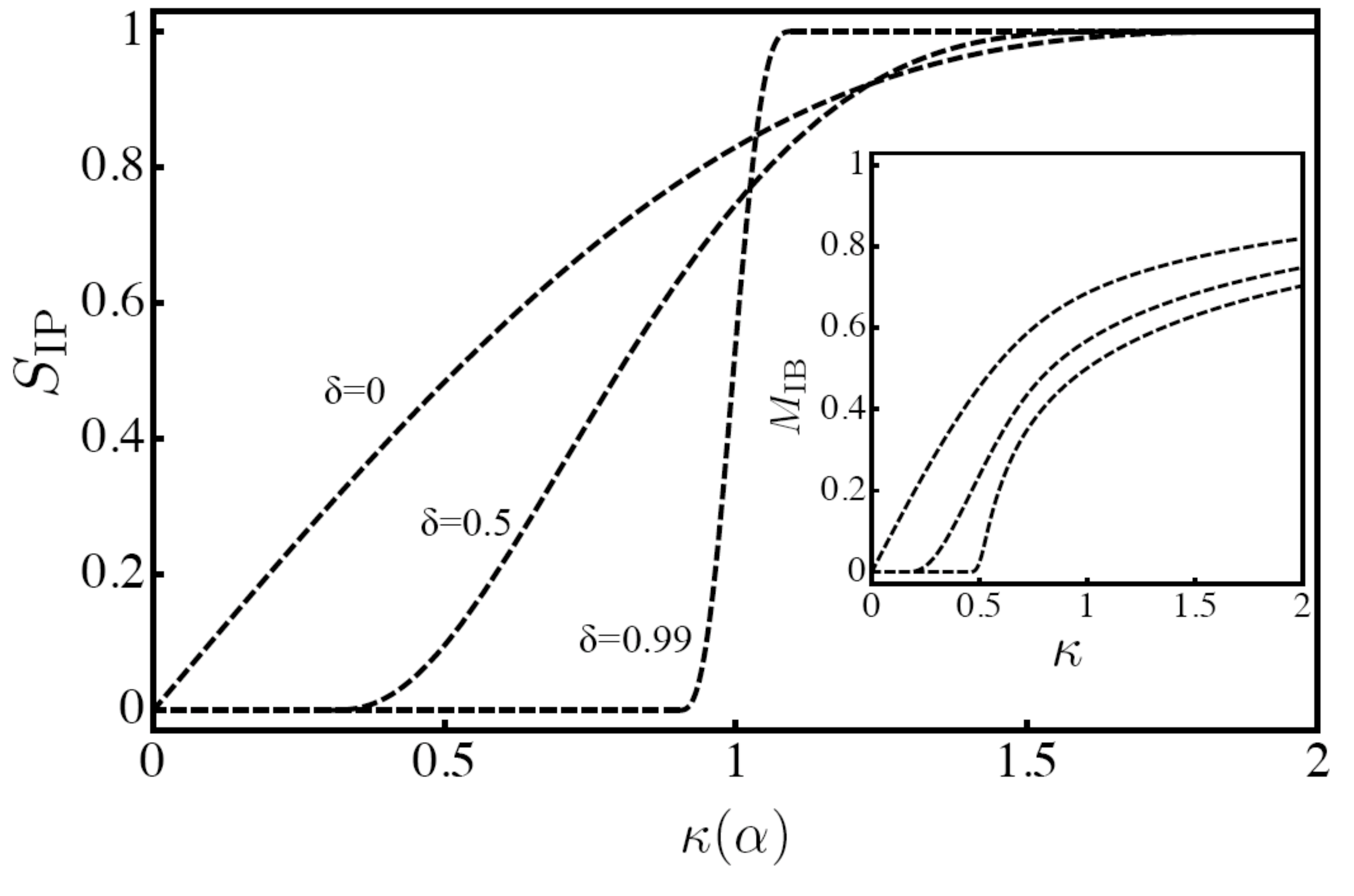}}%
\caption{Zero-temperature scattering rate $S_\mathrm{IP}$ versus kick $\kappa(\alpha)$, for various imbalances: $\delta$ = 0, 0.5, and 0.99. This corresponds to balanced, imbalanced, and nearly polarized. {\bf Inset:} Angle-averaged scattering rate $M_\mathrm{IP}$ versus kick $\kappa$ and at the same three imbalances. Imbalance allows complete suppression to be observed for nonzero $\kappa$.  \label{Fig8}}
\end{center}
\end{figure}

\subsection{Imbalanced Fermi gases \label{sec:imbalance}}

An alternate method of reducing the distribution of initial states is to use the internal structure of the atoms. Consider Raman light scattering between two ground states, and the population of atoms split unequally between them \cite{KetterleImbalance,HuletImbalance} such that $z_i < z_f$, where $i$ and $f$ indicate the initial and final states. Now the thermalized initial and final distributions are 
\begin{equation}
	 n_i(\epsilon)  =  [ 1 + z_i^{-1} \exp{\beta \epsilon} ]^{-1}, \qquad\mbox{and}\qquad
	n_f(\epsilon)  =  [ 1 + z_f^{-1} \exp{\beta \epsilon} ]^{-1}. \label{eq:ninfIP}
\end{equation}
As before, we ignore the change in either distribution due to scattered light or due to interactions. We also assume that incident light excites only atoms from $n_i$ and decays only to $n_f$ \cite{noteRaman}. Integration of (\ref{eq:SSC}) with (\ref{eq:ninfIP}) yields
%
\begin{equation} \label{eq:S3} 
S_\mathrm{IP}(\xi(\alpha), T, \delta) = 1 - \frac{8}{\pi (\beta E_\mathrm{Ff})^3} 
	\int_0^\infty a^{3/2}\, da \int_{-\infty}^\infty dy \,
	\frac{1}{1 + z_i^{-1} e^{a + y^2}} \frac{1}{1 + z_f^{-1} e^{a + (y + \xi(\alpha))^2} },
\end{equation}
where the difference in Fermi energies between the two states is parameterized using $\delta = (E_\mathrm{Ff} - E_\mathrm{Fi})/E_\mathrm{Ff}$. For instance, $N_i = N_f (1-\delta)^3$, etc.  Figures \ref{Fig7} and \ref{Fig8} show the normalized rate of Raman scattering between imbalanced Fermi clouds. At finite temperature, figure~\ref{Fig7} shows $S_\mathrm{IP}$ versus $\delta$ for $\kappa=0.5$, demonstrating that an imbalance enhances suppression. Increasing the cloud imbalance enhances blocking because the range of initial states is increasingly restricted to lower energies. At low temperature, the effect can be dramatic, allowing for {\it complete blocking} at $\delta \geq 3/4$. 

The zero-temperature limit of Raman scattering is 
%
\begin{equation} \fl \label{eq:S3zT}
S_\mathrm{IP}(\kappa(\alpha), \delta) =  \Theta (\kappa(\alpha) +{b}-1) \{ 1-\frac{16}{5 \pi }  \Theta ({b}+{B}) \chi (-{B}/{b})+ \frac{16}{5 \pi } \Theta (1+b-\kappa(\alpha)) \chi (\kappa(\alpha)
   +B)/b^{6} \} ,
\end{equation}
where \( b = \sqrt{1-\delta }  \),  \( B = (\delta -\kappa(\alpha) ^2)/(2 \kappa(\alpha)) \), and $\chi$ is defined in (\ref{eq:chi}). Figure~\ref{Fig8} shows zero-temperature suppression at various imbalances.  In the limit $\delta=0$, there is no imbalance and (\ref{eq:S3zT}) becomes (\ref{eq:S2ZeroT}). The comparison in figure~\ref{Fig8} makes especially clear the wide range of suppression possible at finite recoil momentum for imbalanced gases. Complete suppression is possible even in the angle-averaged case for  $\kappa<0.5$, i.e., for $E_\mathrm{F} > 4 E_\mathrm{R}$. By comparison, in the balanced (or single ground state) case $S_\mathrm{LS} \approx 0.48$ for $\kappa(\alpha)=0.50$. 

In the strong polarization limit $\delta \rightarrow 1$, (\ref{eq:S3zT}) becomes $\Theta(\kappa(\alpha)-1)$. This is reminiscent of the spontaneous emission case (\ref{eq:SEzeroT}), apart from the angular dependence of $\kappa(\alpha)$. We can now see new significance in the results of section~\ref{sec:SE}: in the limit of extreme polarization, the Raman scattering problem is equivalent to the spontaneous emission problem with a two-photon recoil momentum. In both cases, suppression is strong because a second state allows initial states to be exclusively at the middle of the Fermi sea.


Comparing the two enhancement scenarios, the FB approach is less effective than the IP scheme, since initial state selection occurs only along two coordinates instead of all six coordinates.%
\footnote{Another experimentally viable situation would be a cigar-shaped cloud with focused excitation in the plane of symmetry. This would restrict excitation primarily along $q_1$, for instance, but not $q_2$ or $q_3$, and therefore be less effective than the oblate geometry considered in section~\ref{sec:focused}.}
Both schemes are successful at enhancing the expected signal in realistic experimental scenarios: at $\kappa=0.25$ and $T=0.1T_\mathrm{F}$, we find $M_\mathrm{LS}=0.40$, $M_\mathrm{FB}=0.32$ for $\sigma/R_F=0.50$, and $M_\mathrm{IP}=0.07$ for $\delta=0.90$. The dramatic suppression that seemed like a distant experimental prospect in section \ref{sec:SE} (since it required a thermalized excited state atom) is feasible using a Raman light scattering scheme.

\section{Experimental realization and conclusion \label{sec:conclusion}}

To realize a Fermi energy that is $1/\kappa^2$ times the recoil energy, the mean trap strength must be 
\begin{equation} \label{eq:frequ}
\hbar \omega = \frac{E_\mathrm{R}}{\kappa^2 (6 N)^{1/3} }.
\end{equation}
For $10^4$  $^{40}$K atoms, $\omega = 2 \pi \times 860$\,Hz at $E_\mathrm{F}=4 E_\mathrm{R}$. For $10^6$  $^{6}$Li atoms,  $\omega = 2 \pi \times 1620$\,Hz at $E_\mathrm{F}=4 E_\mathrm{R}$.

In order to avoid multiple scattering, the atomic sample must have a low optical density. Consider a cloud of $N$ fermions at $T=0$ in a cylindrically symmetric trap, whose eccentricity is $\epsilon \equiv \omega_z/\omega_\perp$. A Thomas Fermi profile has a resonant optical density bounded by
\begin{equation} \label{eq:OD}
OD^\mathrm{max} = \mathrm{max}\{ \epsilon^{1/3}, \epsilon^{-2/3} \} \frac{3}{4} \left(\frac{E_\mathrm{F}}{E_\mathrm{R}}\right)^2 \frac{E_\mathrm{R}}{\hbar \bar{\omega}}.
\end{equation}
Combining with (\ref{eq:frequ}) and the constraint that $OD < 1$ requires that $N$ be on the order of unity, which is clearly incompatible with a Pauli blocking experiment. Quantitative predictions for spontaneous emission and for light scattering at resonance will require treatment of multiple scattering.

Scattering with off-resonant light can avoid this complication. Detuning approximately $N^{1/6}$ line widths away from resonance can achieve $OD<1$. Our focus on net scattering rates rather than on line shape is partially motivated by this limitation. 

One could use a variety of signatures to search for blocking effects in the lab. 
Because fewer than one photon per atom can be scattered while remaining in the perturbative limit, angle-resolved experiments that measure only a fraction of the total scattered light may be difficult. However, a direct measure of the integrated rate $M$, rather than $S(\alpha)$, is possible through absorption imaging. Another measure of $M$ would be an optical pumping experiment, in which the efficiency of pumping into an occupied Fermi sea is reduced due to blocking effects. In this case the blocking effect would be recorded in atomic populations, circumventing lensing effects of the detuned absorption beam by the degenerate cloud. 


In summary, we have presented two light scattering scenarios in which Pauli blocking is strengthened. Including the effects of inhomogeneous trapping, finite temperature, and without assuming rotational symmetry, we make quantitative predictions for both angle-resolved and angle-averaged signatures. Pauli blocking effects can be enahanced by focusing the excitation beam. Dramatic suppression of incoherent scattering can be achieved with Raman scattering between internal states for large population imbalances and and $E_\mathrm{F} > 4 E_\mathrm{R}$. Our calculations should aid experimental efforts to observe this fundamental quantum optical effect.


\ack {We thank Th.~Busch for additional notes regarding \cite{zoller98}. We also thank B.~DeMarco, D.~F.~V.~James, C.~Salomon, and J.~I. Cirac for stimulating discussion, and the Toronto ultracold atoms group for discussion and proof reading. This work was supported by NSERC and the Canadian Institute for Advanced Research. JHT thanks the MIT-Harvard Center for Ultracold Atoms for their hospitality during the preparation of this manuscript.}

\appendix
\section*{Appendix: Angle-averaged results for zero temperature}
\setcounter{section}{1}

Substitution of (\ref{eq:S2ZeroT}) into (\ref{eq:M}) yields an analytic expression for zero-temperature angle-averaged scattering in the LS case:
%
\begin{equation} \fl \label{eq:M2zTd}
M_\mathrm{LS}(\kappa) = \left\{ \begin{array}{ll}
   \frac{1}{2520 \pi  \kappa ^3} \left\{ 512 (\sqrt{1-\kappa ^2}-1) 
   + 315 \kappa \left(16 \kappa ^2+3\right) \sin ^{-1}(\kappa ) \right.
   & \\
   \left. ~~~~~ + \kappa ^2 \left[ 2 \kappa ^2 \sqrt{1-\kappa ^2} (232 \kappa ^4-1036 \kappa ^2+2157) +2767
   \sqrt{1-\kappa ^2}-3456\right]  \right\}
& \mbox{for $\kappa\leq1$} \\
  1- [ 256 \left(27 \kappa ^2+4\right)-945 \pi  \kappa ] / (5040 \pi  \kappa ^3)
  & \mbox{for $\kappa>1$} \end{array} \right.
\end{equation}
This quantity is plotted as a dashed line in figure~\ref{Fig4}. 

A similar expression can also be found for the FB case  in the tightly focused limit:
\begin{equation} \fl \label{eq:M4zTd}
M^\mathrm{2D}_\mathrm{LS}(\kappa) = \left\{ \begin{array}{ll}
   \frac{1}{210 \pi  \kappa ^3} \left\{  \kappa ^2
   \left[2 \sqrt{1-\kappa ^2} \left(131-32 \kappa ^2\right) \kappa ^2 + 263 \sqrt{1-\kappa
   ^2} - 336\right]
\right. \\ 
\left.
\, + \, 64 \left(\sqrt{1-\kappa ^2}-1\right)+105 \left(4 \kappa ^3+\kappa
   \right) \sin ^{-1}(\kappa)
   \right\} 
& \mbox{for $\kappa\leq1$} \\
1-\frac{21 \kappa  (32 \kappa -5 \pi )+128}{420 \pi  \kappa ^3}  & \mbox{for $\kappa>1$} \end{array} \right.
\end{equation}
The difference between (\ref{eq:M4zTd}) and (\ref{eq:M2zTd}) is plotted as a dashed line in the inset of figure~\ref{Fig6}. 


%
\end{document}